\documentclass[twocolumn]{aastex63}
\usepackage{graphicx}
\usepackage{grffile}

\newcommand{\kmsmpc}{\kms\;{\rm Mpc}^{-1}}

\newcommand{\kms}{{\rm km}\,{\rm s}^{-1}}
\newcommand{\cms}{{\rm cm}^{-2}}
\newcommand{\cmc}{{\rm cm}^{-3}}

\newcommand{\Zsolar}{\;{\rm Z}_{\odot}}
\newcommand{\msolar}{{\rm M}_{\odot}}

\newcommand{\ctssarcmin}{{\rm cts\, s}^{-1}\, {\rm arcmin}^{-2}}
\newcommand{\lxunit}{{\rm erg\; s}^{-1}}

\newcommand{\gad}{{\sc Gadget-3}}

\newcommand{\HI}{{\hbox{H\,{\sc i}}}}

\newcommand{\nh}{{n_{\rm H}}}
\newcommand{\LXextended}{L_{X,>10{\rm kpc}}}

\newcommand{\K}{\,{\rm K}}

\received{}
\revised{}
\accepted{}

\submitjournal{ApJL}

\shorttitle{{\it eROSITA} simulations of the CGM}
\shortauthors{Oppenheimer et al.}

\begin{document}

\title{EAGLE and Illustris-TNG predictions for resolved {\it eROSITA} X-ray observations of the circumgalactic medium around normal galaxies}

\correspondingauthor{Benjamin Oppenheimer}

\author{Benjamin D. Oppenheimer}
\affiliation{CASA, Department of Astrophysical and Planetary Sciences, University of Colorado, 389 UCB, Boulder, CO 80309, USA}
\affiliation{Harvard Smithsonian Center for Astrophysics, 60 Garden Street, Cambridge, MA 02138, USA}

\author{\'{A}kos Bogd\'{a}n}
\affiliation{Harvard Smithsonian Center for Astrophysics, 60 Garden Street, Cambridge, MA 02138, USA}

\author{Robert A. Crain}
\affiliation{Astrophysics Research Institute, Liverpool John Moores University, 146 Brownlow Hill, Liverpool L3 5RF, UK}

\author{John A. ZuHone}
\affiliation{Harvard Smithsonian Center for Astrophysics, 60 Garden Street, Cambridge, MA 02138, USA}

\author{William R. Forman}
\affiliation{Harvard Smithsonian Center for Astrophysics, 60 Garden Street, Cambridge, MA 02138, USA}

\author{Joop Schaye}
\affiliation{Leiden Observatory, Leiden University, P.O. Box 9513, 2300 RA, Leiden, The Netherlands}

\author{Nastasha A. Wijers}
\affiliation{Leiden Observatory, Leiden University, P.O. Box 9513, 2300 RA, Leiden, The Netherlands}

\author{Jonathan J. Davies}
\affiliation{Astrophysics Research Institute, Liverpool John Moores University, 146 Brownlow Hill, Liverpool L3 5RF, UK}

\author{Christine Jones}
\affiliation{Harvard Smithsonian Center for Astrophysics, 60 Garden Street, Cambridge, MA 02138, USA}

\author{Ralph P. Kraft}
\affiliation{Harvard Smithsonian Center for Astrophysics, 60 Garden Street, Cambridge, MA 02138, USA}

\author{Vittorio Ghirardini}
\affiliation{Max-Planck-Institut f{\"u}r extraterrestrische Physik, Giessenbachstra{\ss}e, 85748 Garching, Germany}
\affiliation{Harvard Smithsonian Center for Astrophysics, 60 Garden Street, Cambridge, MA 02138, USA}

%\affiliation{CASA CU Boulder}

\email{benjamin.oppenheimer@colorado.edu}

\begin{abstract}

We simulate stacked observations of nearby hot X-ray coronae associated with galaxies in the EAGLE and Illustris-TNG hydrodynamic simulations.  A forward modeling pipeline is developed to predict 4-year {\it eROSITA} observations and stacked image analysis, including the effects of instrumental and astrophysical backgrounds.  We propose an experiment to stack $z\approx 0.01$ galaxies separated by specific star-formation rate (sSFR) to examine how the hot ($T\geq 10^6$ K) circumgalactic medium (CGM) differs for high- and low-sSFR galaxies.  The simulations indicate that the hot CGM of low-mass ($M_*\approx 10^{10.5}\ \msolar$), high-sSFR (defined as the top one-third ranked by sSFR) central galaxies will be detectable to a galactocentric radius $r \approx 30-50$ kpc.  Both simulations predict lower luminosities at fixed stellar mass for the low-sSFR galaxies (the lower third of sSFR) with Illustris-TNG predicting $3\times$ brighter coronae around high-sSFR galaxies than EAGLE. Both simulations predict detectable emission out to $r \approx 150-200$ kpc for stacks centered on high-mass ($M_*\approx 10^{11.0}\ \msolar$) galaxies, with EAGLE predicting brighter X-ray halos.  The extended soft X-ray luminosity correlates strongly and positively with the mass of circumgalactic gas within the virial radius ($f_{\rm CGM}$). Prior analyses of both simulations have established that $f_{\rm CGM}$ is reduced by expulsive feedback driven mainly by black hole growth, which quenches galaxy growth by inhibiting replenishment of the ISM.  Both simulations predict that {\it eROSITA} stacks should not only conclusively detect and resolve the hot CGM around $L^*$ galaxies for the first time, but provide a powerful probe of how the baryon cycle operates, for which there remains an absence of consensus between state-of-the-art simulations.  

\end{abstract}

% Galaxy accretion (575)

\section{Introduction} \label{sec:intro}

Extended hot X-ray coronae have long been theorized to supply the gas necessary for star-formation in disc galaxies  \citep{spitzer56, white78}.  \citet{white91} predicted emission levels that should have been readily detected by {\it Chandra} and {\it XMM-Newton}.  The initial surprise of weak or no detection of soft X-ray emission from disc galaxies \citep[e.g.][]{benson00,li06} has been interpreted as a signature of superwind feedback removing gas from halos and leaving behind substantially flattened central hot gas profiles \citep{crain10}.

Pointed observations have revealed X-ray coronae associated with individual, isolated elliptical galaxies \citep[e.g.][]{forman85, osullivan01, goulding16}, whilst stacking {\it ROSAT} all-sky survey data about the coordinates of mainly early-type galaxies has revealed a strong correlation between the inferred CGM mass fraction and galaxy mass \citep{anderson15}. Detections associated with individual disc galaxies are primarily limited to rare, massive cases \citep[e.g.][]{bogdan13a, bogdan13b, li17}.  

The {\it eROSITA} instrument on the Spectrum-Roentgen-Gamma mission \citep{erosita} launched in July 2019 will map the entire sky at $30\times$ greater sensitivity and higher spatial resolution than {\it ROSAT}, opening new possibilities to not only detect, but also resolve, the structure of emission from galaxies Milky Way-mass and below.  Tenuous, diffuse X-ray halos around $L^*$ galaxies ($M_*\ga 10^{10}\ \msolar$) are a ubiquitous prediction of realistic cosmological hydrodynamical simulations, including the EAGLE \citep{schaye15,crain15,mcalpine16} and Illustris-TNG \citep[][hereafter TNG]{pillepich18,nelson18a} simulations. 

Both these simulations broadly reproduce fundamental galaxy properties, including stellar mass functions, passive galaxy fractions, and morphological types in $\sim 100^3$ Mpc$^3$ hydrodynamic volumes containing thousands of $L^*$ galaxies.  However, EAGLE and TNG apply distinct prescriptions for stellar and super-massive black hole (SMBH) feedback that result in markedly different CGM masses at $z=0$ \citep[][hereafter D20]{davies20}.  The feedback energy imparted over cosmic time is often enough to unbind a significant fraction of the CGM beyond the virial radius \citep[][D20]{opp20}.  The notable differences in how energetic feedback operates as a function of galaxy type between EAGLE and TNG should make divergent and \textit{testable} predictions for observations by X-ray telescopes with large collecting areas \citep{davies19, truong20}. 

Observational characterization of the CGM has to date been driven primarily by UV absorption line observations of $\HI$ and metal ions in sightlines intersecting the gaseous environments of galaxies \citep[e.g.][]{tumlinson11,stocke13,liang14,turner14}.  These UV species mainly trace $T=10^{4-5.5}$ K gas \citep[e.g.][]{ford13,rahmati16}, with diffuse metals indicating the presence of heavy elements transported from the ISM by superwind feedback \citep[e.g.][]{aguirre01,opp16,nelson18b}. The total mass of the UV-traced CGM appears to be greater than that of the central galaxy \citep{werk14,prochaska17}, but simulations predict hot ($T\geq 10^6$ K) CGM masses that further outweigh the $T<10^6$ K CGM, even for $L^*$ disc galaxies \citep{ford14,opp18c}. The hot CGM component therefore has the potential to prove more constraining for the total gaseous content of galactic halos. Additionally, the hot component almost certainly contains the vast majority of the CGM energy, which, if measured, would provide essential constraints on the ultimate fate of momentum and entropy from feedback.  

The {\it eROSITA} mission will average a 2 ksec integration upon the release of its final all-sky survey (eRASS:8) comprising 4 years of observations.  This letter makes {\it eROSITA} stacking predictions for nearby galaxies from the EAGLE and TNG simulations.  The 15" spatial resolution of {\it eROSITA} should allow interior X-ray profiles to be resolved for nearby halos, which is why we propose stacking galaxies at $z\approx 0.01$. 

In \S\ref{sec:methods}, we introduce the EAGLE and TNG simulations and our forward modeling technique to predict results from stacked {\it eROSITA} observations.  We present the main results in \S\ref{sec:results} and discuss their interpretation in \S\ref{sec:discuss}.  We summarize in \S\ref{sec:summary}.  We use a cosmology of $\Omega_{\rm M}=0.307$, $\Omega_{\Lambda}=0.693$, $H_{0}= 67.77\ \kmsmpc$, and $\Omega_{\rm b}=0.04825$ for our mock observations.  

\section{Methods} \label{sec:methods}

\subsection{Simulations} 

The EAGLE Ref-L100N1504 simulation \citep{schaye15} is a $100^3$ Mpc$^3$ smoothed particle hydrodynamics (SPH) run with a modified version of the \gad~code \citep{springel05} using a pressure-entropy implementation of SPH.  It uses $1504^3$ SPH and dark matter (DM) particles.  The TNG-100 simulation \citep{pillepich18} uses the AREPO \citep{springel10} moving mesh hydro solver in a volume of $110^3$ Mpc$^3$ with $1820^3$ DM particles and initial gas cells.  Both simulations have $\sim 1$ kpc gravitational softening lengths and gas and stellar mass resolutions of $\sim 10^6\ \msolar$.  The two models incorporate significantly different subgrid prescriptions for stellar and AGN feedback, but in both cases the relevant parameters were calibrated to ensure the reproduction of key observables. For EAGLE, these were present-day stellar masses ($M_*$), galaxy disc sizes, and SMBH masses ($M_{\rm SMBH}$).  For TNG, the cosmic star formation history, galaxy star formation rates (SFR), and the gas fractions of galaxy groups were also considered. We use the $z=0$ snapshot that contains $2199$ ($3808$) central galaxies with $M_*>10^{10}\ \msolar$ for EAGLE (TNG).  
%XXX- \citep{schaller15} taken out.  

\subsection{Simulation galaxy samples}

To make observationally-reproducible samples, we select simulated central galaxies based on $M_*$ and sSFR $\equiv {\rm SFR}/M_*$.  We define two stellar mass bins called ``low-mass'' and ``high-mass.''  We use the EAGLE simulation to define the mass ranges, such that the low-mass (high-mass) bin spans centrals from $M_*=10^{10.2-10.7}$ ($10^{10.7-11.2}$) $\msolar$.  Stellar mass limits can be converted to volume densities by rank-ordering central $M_*$ and selecting the volume density for galaxies greater than a given $M_*$ (e.g. $M_*>10^{10.2}\ \msolar$ in EAGLE corresponds to $1.60\times 10^{-3}$ Mpc$^{-3}$).  Hence, the low-mass (high-mass) limits correspond to volume densities of $1.60\times 10^{-3}-5.13\times10^{-4}$ ($5.13\times 10^{-4}-5.6\times10^{-5}$) Mpc$^{-3}$.  To select TNG galaxies with the same volume density, we need to use appreciably higher mass limits, because TNG has $0.1-0.2$ dex higher average stellar masses than EAGLE at $M_*=10^{10.4-11.4}\ \msolar$.  The TNG low- and high-mass bins are $10^{10.38-10.82}$ and $10^{10.82-11.39}\ \msolar$.  By normalizing to volume density in two simulations that use nearly identical cosmologies, the halo masses are similar across the two simulations for each bin (Table \ref{tab:samples}).  

\begin{table}
\caption{Simulation galaxy counts in mock samples}
\begin{center}
\begin{tabular}{lcc}
\hline
\hline
EAGLE & Low-Mass, $M_*=$ & High-Mass, $M_*=$\\  
      & $10^{10.20-10.70}\ \msolar$ &  $10^{10.70-11.20}\ \msolar$ \\
\hline 
\# High-sSFR & 357 & 144 \\ 
sSFR Threshold & $\geq 10^{-10.26}$ yr$^{-1}$ &  $\geq 10^{-10.51}$ yr$^{-1}$ \\
$M_{200}$ Range$^1$ & $10^{11.92-12.30}\ \msolar$ & $10^{12.37-12.87}\ \msolar$ \\
\hline
\# Low-sSFR & 356 & 143 \\ 
sSFR Threshold & $< 10^{-10.67}$ yr$^{-1}$ &  $<10^{-11.54}$ yr$^{-1}$ \\
$M_{200}$ Range & $10^{11.97-12.45}\ \msolar$ & $10^{12.50-13.01}\ \msolar$ \\
\hline
\hline
Illustris-TNG & Low-Mass, $M_*=$ & High-Mass, $M_*=$ \\  
      & $10^{10.38-10.82}\ \msolar$ &  $10^{10.82-11.39}\ \msolar$ \\
\hline
\# High-sSFR & 482 & 200 \\ 
sSFR Threshold & $\geq 10^{-10.21}$ yr$^{-1}$ &  $\geq10^{-11.14}$ yr$^{-1}$ \\
$M_{200}$ Range & $10^{11.91-12.19}\ \msolar$ & $10^{12.31-12.98}\ \msolar$ \\
\hline
\# Low-sSFR & 481 & 199 \\ 
sSFR Threshold & $< 10^{-12.38}$ yr$^{-1}$ &  $<10^{-12.60}$ yr$^{-1}$ \\
$M_{200}$ Range & $10^{11.99-12.43}\ \msolar$ & $10^{12.41-12.97}\ \msolar$ \\
\hline
\end{tabular}
\end{center}
\parbox{25cm}{
$^1$ $1\sigma$ range for $M_{200}$ values in sample.
}
\label{tab:samples}
\end{table}

%EAGLE       Low-SSFR  High-SSFR
%Low-Mass   12.21+0.24-0.24    12.07+0.23-0.15
%High-Mass   12.70+0.31-0.20  12.60+0.27-0.23

%TNG
%Low-Mass   12.18+0.25-0.19   12.03+0.16-0.12
%High-Mass   12.66+0.31-0.25   12.63+0.35-0.32

We create samples divided into bins of sSFR and define high and low-sSFR as the upper and lower thirds of the sSFR distribution.  The resulting sSFR thresholds for each $M_*$ bin are listed in Table \ref{tab:samples}.  The main motivation for these samples is that D20 showed that, for both EAGLE and TNG, sSFR is highly correlated with the gas content of the CGM, defined as
\begin{equation} \label{equ:fCGM}
f_{\rm CGM} \equiv \frac{M_{\rm gas}(R<R_{200})}{M_{200}(R<R_{200})} \times \frac{\Omega_{\rm M}}{\Omega_{\rm b}},
\end{equation}
where $R_{200}$ and $M_{200}$ are respectively the radius and mass of the sphere, centered on a galaxy, with mean enclosed density of $200\rho_{\rm crit}$, and $\rho_{\rm crit}$ is the critical density for closure. A key objective of this stacking exercise will be to assess whether the diffuse X-ray luminosity of galaxies (at fixed $M_*$) indeed correlates with $f_{\rm CGM}$.

TNG has 35\% more volume than EAGLE, and therefore a larger sample size for fixed $M_*$.  TNG has a wider range of sSFR values resulting in a larger gap in sSFR thresholds.  Our aim is to design an experiment where observers can create samples of galaxies ranked by sSFR, without reliance on matching absolute values.  For brevity, the intermediate sSFR bin is not discussed.  We exclude halos with $M_{200}>10^{13.3}\ \msolar$, which only affects high-mass samples, because our mocks indicate these X-ray halos are individually detectable by {\it eROSITA}. 

We stack 100 (50) low-mass (high-mass) galaxies at a time, observed at $z=0.01$.  Based on volume densities for both sSFR bins, we expect 230 (95) low-mass (high-mass) galaxies per bin to be located at an average distance of $z=0.01$ for the entire sky with galactic latitude $\mid b\mid >15^\circ$.  We make conservatively small samples given that ground-based surveys and/or data releases may access less than half of the sky.  Our goal is to create the nearest sample, limited by the volume of the local Universe, that can be used to stack and spatially resolve extended emission.  Galaxies in the real Universe will reside at a variety of distances, but our $z=0.01$ stacks are representative of local galaxies where contamination from galactic sources (X-ray binaries, hot ISM) should be mostly limited to the inner $r=1'$ ($12$ kpc at $z=0.01$).  We tested stacking thousands of $z=0.03$ galaxies, finding similar results but with a reduced ability to resolve the emission structure for $r \lesssim 30$ kpc.  

\begin{figure*}
\includegraphics[width=1.00\textwidth]{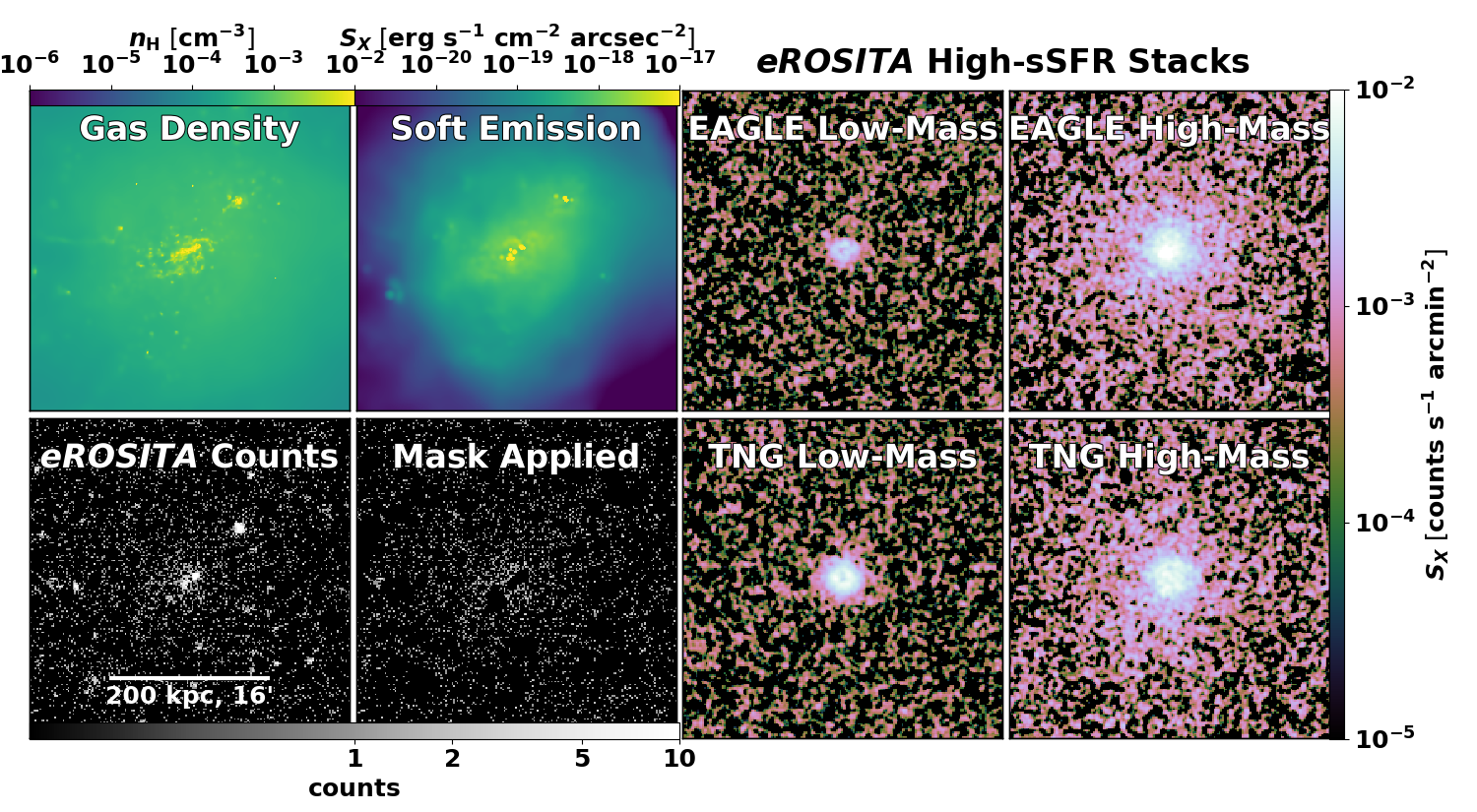}
\caption{{\it Left 4 panels:} An EAGLE $M_{200}=10^{12.58}\ \msolar$ halo hosting $M_*=10^{10.73}\ \msolar$ star-forming, late-type galaxy.  The density (upper left) and soft (0.5-2.0 keV) X-ray emissivity (upper right) are shown in $400\times400$ kpc snapshot images.  A mock $z=0.01$ {\it eROSITA} count map is generated (lower left) and point source-like objects are masked, including CXB sources, the prominent satellite in the upper right, and emission on top of the galaxy, leaving behind an extended halo (lower right).  This halo is brighter than typical, $\LXextended=10^{41} \lxunit$, and most halos do not show individually detectable emission. {\it Right 4 panels:} Mock {\it eROSITA} stacks of high-sSFR galaxies, including stacks of 100 low-mass galaxies (left panels) and 50 high-mass galaxies (right panels) for EAGLE (upper panels) and TNG (lower panels).  These panels also span $400\times 400$ kpc.  }
%Figure made with APLpy \citep{aplpy}.}
\label{fig:pipeline}
\end{figure*}

\subsection{Forward Modeling Pipeline}

We use the pyXSIM package\footnote{\url{http://hea-www.cfa.harvard.edu/~jzuhone/pyxsim/}} \citep{zuhone16} to create a SIMPUT\footnote{\url{http://hea-www.harvard.edu/heasarc/formats/simput-1.1.0.pdf}} file of mock photons emanating from hot, diffuse plasma out to $3R_{200}$ for each halo.  An example EAGLE halo is shown in the left four panels of Fig. \ref{fig:pipeline}.  For each fluid element with $T>10^{5.3}\K$ and hydrogen number density $\nh<0.22~\cmc$ within this region, pyXSIM randomly generates photons using a Monte-Carlo sampling of X-ray spectra from the Astrophysical Plasma Emission Code \citep[{\sc APEC};][]{smith01}.  {\sc APEC} spectra assume collisional ionization equilibrium given the density, temperature, and metallicity (including 9 individually-tracked abundances) of each fluid element.  We do not simulate X-rays from the ISM. 
%, corresponding to densities $\ga 0.1\ \cmc$.  

In addition to the source photons, we include simulated Galactic foreground emission and a Cosmic X-ray background (CXB) randomly-generated using the SOXS package\footnote{\url{http://hea-www.cfa.harvard.edu/~jzuhone/soxs/}; background described in \url{http://hea-www.cfa.harvard.edu/~jzuhone/soxs/users_guide/background.html}}.  Galactic absorption assuming a column of $N_{\rm HI}=2\times10^{20} \cms$ is applied to the source and CXB photons.  

The SIXTE simulation software \citep{sixte} uses SIMPUT file inputs to create {\it eROSITA} 2 ksec observations centered at the position of the galaxy.  The {\tt erosim} tool generates event files for the seven {\it eROSITA} cameras and combines them into one image, as shown in Fig.~\ref{fig:pipeline} (lower left panel) with energy clipped to show only soft X-ray counts ($0.5-2.0$ keV).  The image, which includes the instrumental background and point spread function, is dominated by CXB photons. %, while the source photons are sub-dominant.  

The CIAO \citep{ciao} {\tt wavdetect} routine detects concentrated sources, including CXB sources, bright satellites, and point source-like emission at the position of the galaxy, which we then mask.  Given that we do not include galactic ISM nor expected contributions from X-ray binaries, which should dominate at the position of the stellar component, we focus on emission outside a projected radius $r>10$ kpc at $z=0.01$.  Individual masked images with $9.6''$ pixels are added together in our mock stacks, as are the individual exposure maps that include the {\tt wavdetect}-generated masks.  We make an off-source stack using the same procedure performed without galaxy halo emission.  Both stacks are divided by their respective summed exposure maps to obtain $\ctssarcmin$, and the off-source stack is subtracted from the on-source stack.  Four reduced $z=0.01$ stacks of high-sSFR galaxies, low-mass and high-mass samples for EAGLE and TNG, are shown in Fig. \ref{fig:pipeline} (right panels).

\section{Results} \label{sec:results}

\begin{figure*}
\includegraphics[width=0.50\textwidth]{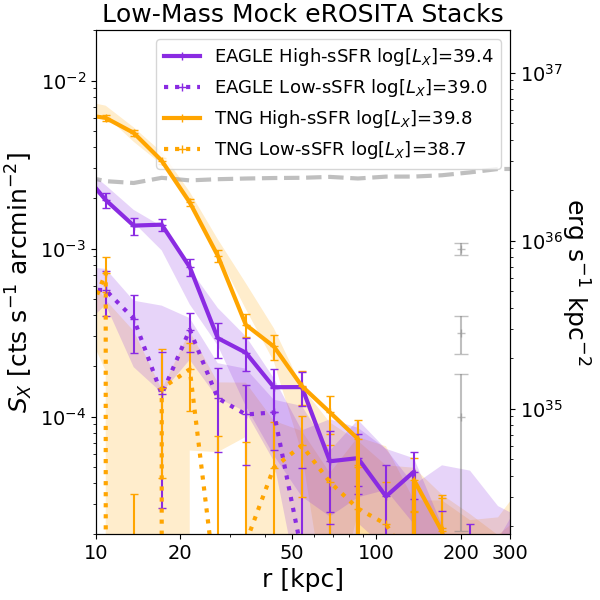}
\includegraphics[width=0.50\textwidth]{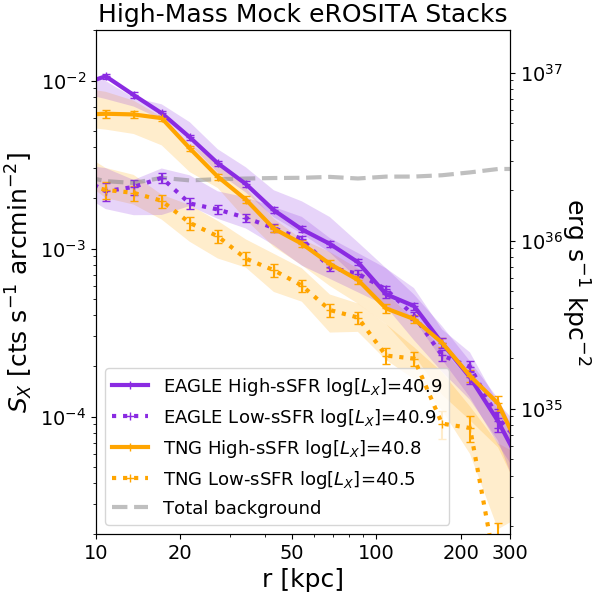}
\caption{Simulated {\it eROSITA} 4-year of mean soft X-ray surface brightness profiles around low-mass (left) and high-mass (right) halos in EAGLE and TNG.  Colored lines indicate one mock survey of 100 low-mass and 50 high-mass $z=0.01$ galaxy stacks with Poisson error bars, which should be reproducible with half the all-sky survey.  Shading indicates $1\sigma$ spreads from 20 mock surveys.  Average $\LXextended/\lxunit$ values calculated from the stacks are listed in the legend.  Both simulations predict brighter X-ray halos around higher sSFR galaxies.  TNG predicts a greater dichotomy at low mass, and EAGLE predicts brighter halos overall at high mass.  {\it eROSITA} should enable detection of star-forming galaxy halos out to $30-50$ kpc around low-mass galaxies and of all halos out to $150-200$ kpc around  high-mass galaxies.  The total astrophysical and instrumental background is indicated by the gray dashed lines.  We plot example error bars in gray indicating $3\%$ of the background level in the left panel to demonstrate the effect of possible systematic errors.  }
\label{fig:SB_profiles}
\end{figure*}

Figure \ref{fig:SB_profiles} shows radial soft X-ray ($0.5-2.0$ keV) surface brightness profiles ($S_{X}$) for the four samples of low-mass (left panel) and high-mass (right panel) galaxies.
Purple (orange) lines show EAGLE (TNG) simulations for high-sSFR (solid) and low-sSFR (dotted) samples.  There are 100 low-mass and 50 high-mass galaxies in each mock survey, shown along with Poisson error bars.  Shaded regions correspond to $1\sigma$ spreads of 20 mock surveys.  We calculate and list the average extended soft X-ray luminosity, $\LXextended$, by integrating $S_{X}$ between 10 and 200 kpc, and converting to $\lxunit$ using {\it eROSITA}'s area response function with an average collecting area of $2100$ cm$^{2}$ and a mean photon energy of $0.8$ keV that we obtain from our SIMPUT files.

Most $z=0.01$ low-mass stacks appear to be detectable out to 50 kpc at a level of $10^{-4}~\ctssarcmin$.  TNG predicts high-sSFR galaxies to be brighter in the inner 50 kpc and have $15\times$ higher luminosities than low-sSFR galaxies.  EAGLE predicts a similar trend, but a much smaller difference of $2.5\times$.  The coronae of low-mass, high-sSFR galaxies are 3$\times$ brighter in TNG than in EAGLE.  All high-mass stacks appear to be detectable out to $r \ga 200$ kpc with EAGLE predicting more luminous X-ray halos.  Both simulations predict stronger interior ($r<30$ kpc) emission around high-sSFR galaxies, but EAGLE predicts very similar profiles at $r>50$ kpc in contrast to TNG, which predicts stronger emission around high-sSFR galaxies everywhere.  

The detection of extended hot halos relies on stable subtraction of the background, which has a level of $2.5-3.0\times 10^{-3}~\ctssarcmin$ and is indicated by the gray dashed line.  Our pipeline suggests that it should be possible to detect count rates at up to $30\times$ below the background, which agrees well with the predicted background calculated in the {\it eROSITA} Science Booklet \citep{erosita}. The error bars in Fig. \ref{fig:SB_profiles} indicate only Poisson errors from the source and background stacks added in quadrature.  The shaded regions represent an estimate of cosmic variance when stacking the galaxies contained within the simulation volumes, which can exceed Poisson errors, especially for low-mass, low-sSFR stacks. 

Systematic errors that are not included in our pipeline may make it more difficult to detect source counts as low as 3\% of the background level.  We demonstrate the size of systematic errors using three stand-alone error bars in the left panel if we assume systematic errors of 3\% the background, which is a precedent expected from previous {\it Chandra} and {\it XMM-Newton} data processing.  This would raise the detectability threshold to $\sim 2\times 10^{-4}~\ctssarcmin$, reducing the maximum radius out to where we can detect low-mass (high-mass), high-sSFR stacked emission to $\sim 30$ ($\sim 150$) kpc.  

Smaller stacks including fewer galaxies should be able to test these models.  We predict stacking only 30 low-mass galaxies will distinguish TNG high- and low-sSFR stacks, as well as EAGLE and TNG high-sSFR stacks from each other.  This also means that our proposed experiment could bear similar results with less total integration time using 100 low-mass galaxies, perhaps as soon as the eRASS:2 ($1$-year) data release.  

We also perform a test where we take the median stacked $S_{X}$ instead of the mean, finding the same results, including integrated $\LXextended$ values, within 0.2 dex.  This indicates extended emission is smooth, because discrete sources would create a patchy distribution and much lower medians relative to means.  

\begin{figure}
\includegraphics[width=0.46\textwidth]{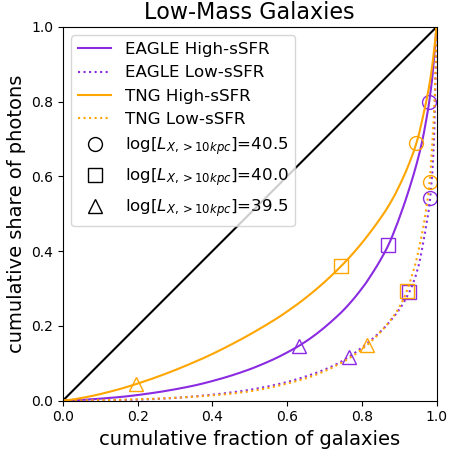}
\caption{Low-mass $z=0.01$ galaxy samples are rank-ordered by soft X-ray photon counts outside $r=10$ kpc to demonstrate the relative share of extended emission arising from different galaxies within each stack.  The black line demonstrates a completely equal distribution.  High-sSFR X-ray halos are distributed more uniformly than low-sSFR halos.  Symbols indicate the fraction of galaxies with extended luminosities fainter than the values listed in the legend.  Deep {\it eROSITA} observations of individual halos will be able to complement stacking observations by constraining the upper portions of these curves.}\label{fig:Gini}
\end{figure}

The eRASS:8 scanning pattern will provide deeper coverage at the ecliptic poles with 550 deg$^2$ scanned at $>10$ ksec; therefore we offer predictions for the distributions of individual halo emission that {\it eROSITA} should be able to probe in these deeper regions.  We rank order halos by extended emission outside $r=10$ kpc ($50"$ at $z=0.01$) in each low-mass sample, and plot the cumulative photon contribution in Figure \ref{fig:Gini}.  The brightest low-mass stack, TNG high-sSFR galaxies, is also the most uniformly distributed, but nonetheless both simulations predict that low-sSFR galaxies are much more dominated by outliers than their high-sSFR counterparts.  We quantify the inequality of $S_{X}$ using the Gini statistic
\begin{equation}
G_{S_{X}} = \frac{\sum\limits_{i=1}^{n}\sum\limits_{j=1}^{n} \mid L_{X,>10 {\rm kpc},i}-L_{X,>10 {\rm kpc},j} \mid}{2 n^2 {\bar{L}_{X,>10 {\rm kpc}}}},
\end{equation}
where $\bar{L}_{X,>10{\rm kpc}}$ is the mean extended luminosity of $n$ galaxies.  We report $G_{S_{X}}$, which is twice the geometric area between the locus of equality (solid black line) and each colored curve in Fig. \ref{fig:Gini}.  For EAGLE (TNG), high-sSFR galaxies $G_{S_{X}} = 0.66$ ($0.51$), and for low-sSFR galaxies $G_{S_{X}} = 0.83$ ($0.83$).  Open symbols show the fraction of galaxies that have CGM luminosities smaller than the corresponding value indicated in the legend.  For example, open squares show that 58\% (64\%) of extended emission comes from the 13\% (26\%) of brightest high-sSFR low-mass halos with $L_{X,>10{\rm kpc}}\geq 10^{40.0} \lxunit$ in EAGLE (TNG).  Low-sSFR halos have more diversity in $M_{200}$, which results in the brightest halos dominating the low-mass stacks.  High-mass galaxies have $G_{S_{X}}=0.60-0.73$.   

We also experiment using fixed stellar mass bins, because X-ray emission correlates with sSFR and the integrated star formation is of course encoded in $M_*$.  Unsurprisingly, EAGLE (TNG) luminosities increase (decrease) by $0.1-0.2$ dex, owing to EAGLE stellar masses increasing relative to TNG compared to the normalized volume density samples.  Halo masses are higher for EAGLE than TNG when using fixed $M_*$ bins, while they are mainly overlapping for the normalized volume density samples (see Table \ref{tab:samples} for $M_{200}$ mass ranges).  

\section{Discussion} \label{sec:discuss}

Other publications have compared these simulations to existing X-ray observations of similar systems.  \citet[][their Appendix A]{davies19} show EAGLE $L_{X}$ values are in the range of individually observed objects at $M_{200}\la 10^{13}\ \msolar$, but the extended emission around more massive halos in EAGLE (mostly excluded in our samples) is too bright \citep[see also][]{schaye15}.  \citet{truong20} showed TNG predicts $\sim 10\times$ greater emission from galaxies at fixed mass with blue colors than red colors, which is consistent with our low-mass sSFR-divided samples; however, their approach is quite different to ours, as they concentrate on emission within the half-light radius of the galaxy and exclude the faint X-ray halos that are our focus.  The \citet{truong20} low-mass star-forming central luminosities appear to be brighter than the late-types observed by \citet{li13}, but it remains to be seen how much of a discrepancy this is and how the extended emission compares.  

Existing deep imaging of a handful of individual X-ray halos is capable of probing the $S_{X}$ values of our stacks, as in the cases of massive spirals (NGC 1961, \citep{anderson16}; NGC 6753, \citep{bogdan17} and the sample of \citet{li17}). The emission at $r\approx 50$ kpc from these halos is several times less than that of the high-mass high-sSFR stacks from both simulations.  These galaxies were targeted based on being X-ray-bright and massive late-types.  If they are representative of the galaxies in our {\it eROSITA} simulated stacks, the observations suggest that both EAGLE and TNG over-predict extended emission from high-mass star-forming galaxies in general.  The hot gas fractions of galaxy groups in EAGLE are known to be too high \citep{schaye15}, and it is plausible that the expulsion of gas from galaxy-scale halos is also too weak.  We also find that the metallicity of the central hot CGM of EAGLE galaxies is generally higher than the $\sim 0.1 \Zsolar$ derived for NGC 1961 and NGC 6753.  However, there is the possibility that these selected galaxies are not wholly representative of the local volume-selected sample (without regard to galaxy type) presented above.  Hence, in the absence of extensive additional {\it XMM-Newton} or {\it Chandra} observations, only the proposed {\it eROSITA} dataset can provide definitive constraints. If {\it eROSITA} observes fainter stacked emission than either EAGLE or TNG, then future simulations of the galaxy population will need to ensure that in addition to reproducing key stellar properties of galaxies, the implementation and calibration of their feedback implementations satisfies these complementary constraints.

%\begin{figure*}
%\includegraphics[width=0.50\textwidth]{figures/fcgm_LXextended.EAGLE_Low-Mass.cat.nparam2.png}
%\includegraphics[width=0.50\textwidth]{figures/fcgm_LXextended.EAGLE_High-Mass.cat.nparam2.png}
%\includegraphics[width=0.50\textwidth]{figures/fcgm_LXextended.TNG_Low-Mass.cat.nparam2.png}
%\includegraphics[width=0.50\textwidth]{figures/fcgm_LXextended.TNG_High-Mass.cat.nparam2.png}
%\caption{Extended ($r>10$ kpc) X-ray emission as a function of CGM mass fraction ($f_{\rm CGM}$) divided into high and low-sSFR galaxies for low-mass (left panels) and high-mass (right panels) EAGLE (upper panels) and TNG (lower panels) halos.  Median and $1\sigma$ spreads are indicated by large points with error bars.  High-sSFR galaxies reside in halos with higher gas fractions than low-sSFR galaxies, with this trend being most pronounced for low-mass TNG galaxies.  Extended X-ray luminosity is strongly correlated with $f_{\rm CGM}$ and is also dependent on $M_{200}$ as indicated by symbol size.  $f_{\rm CGM}$ equals unity for a halo that retains the cosmic proportion of baryons entirely in the CGM (see Eq. \ref{equ:fCGM}).  }
%\label{fig:fgas_LX}
%\end{figure*}

\begin{figure*}
\includegraphics[width=0.50\textwidth]{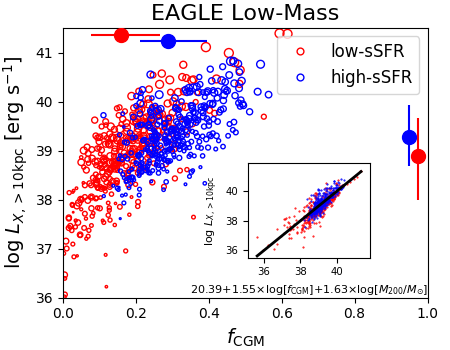}
\includegraphics[width=0.50\textwidth]{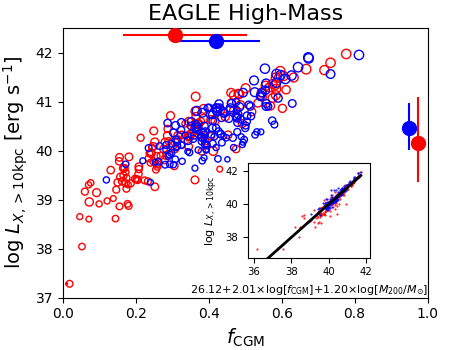}
\includegraphics[width=0.50\textwidth]{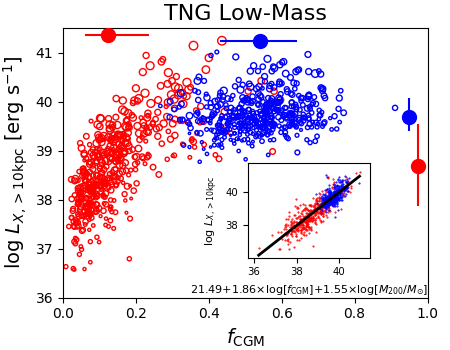}
\includegraphics[width=0.50\textwidth]{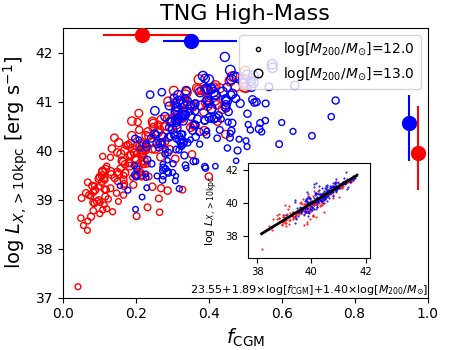}
\caption{Extended ($r>10$ kpc) X-ray emission as a function of CGM mass fraction ($f_{\rm CGM}$) divided into high and low-sSFR galaxies for low-mass (left panels) and high-mass (right panels) EAGLE (upper panels) and TNG (lower panels) halos.  Median and $1\sigma$ spreads are indicated by large points with error bars.  High-sSFR galaxies reside in halos with higher gas fractions than low-sSFR galaxies, with this trend being most pronounced for low-mass TNG galaxies.  Extended X-ray luminosity is strongly correlated with $f_{\rm CGM}$ and is also dependent on $M_{200}$ as indicated by symbol size.  $f_{\rm CGM}$ equals unity for a halo that retains the cosmic proportion of baryons entirely in the CGM (see Eq. \ref{equ:fCGM}).  Inset panels show two-parameter linear regressions indicating the combinations of log[$f_{\rm CGM}$] and log[$M_{200}/\msolar$] that best reproduce log[$\LXextended$] (equations below inset $x$-axes, black lines represent fits). }
\label{fig:fgas_LX}
\end{figure*}

\subsection{X-ray emission traces CGM baryon content}

While soft X-ray emission around $L^*$ galaxies is strongly biased to the densest gas and is dominated by metal-line emission \citep{crain13}, \citet{davies19} showed that $L_{X}$ is highly correlated with the total CGM gas content in EAGLE.  We show the extended X-ray luminosity as a function of $f_{\rm CGM}$ in Fig. \ref{fig:fgas_LX}. Medians and $1\sigma$ spreads are indicated along the top ($f_{\rm CGM}$) and to the right ($L_{X,>10{\rm kpc}}$).   

We propose that extended emission in {\it eROSITA} stacks provides an effective proxy for CGM baryon content.  The low-mass TNG bin has the largest difference between high and low-sSFR $f_{\rm CGM}$ values ($0.54$ vs. $0.12$; D20), which primarily drives the remarkable prediction from TNG that high-sSFR galaxies should have $15\times$ greater coronal X-ray luminosity than low-sSFR galaxies of the same $M_*$.  The difference in $\LXextended$ is only a factor of $2.5$ for EAGLE, which reflects the narrower range of $f_{\rm CGM}$ (medians of $0.29$ vs. $0.16$ for these samples).  

Typical halo masses in the low-mass stacks are $M_{200}\approx 10^{12.0-12.3}\ \msolar$ with low-sSFR galaxies having a median halo mass $0.15$ dex higher than high-sSFR galaxies in both simulations.  $\LXextended$ at fixed $f_{\rm CGM}$ is higher for more massive halos, where $M_{200}$ is denoted by the symbol size in Fig. \ref{fig:fgas_LX}.  The high-mass stacks exhibit the same overall behavior, but with galaxies occupying more massive halos ($M_{200}\approx 10^{12.4-13.0}\ \msolar$).  There is also less difference between the high and low-sSFR subsamples and less scatter.  

We perform linear regressions to produce least squares fits to $\LXextended$ using $f_{\rm CGM}$ and $M_{200}$ in logarithmic space, and plot the results in inset panels of Fig. \ref{fig:fgas_LX} with the best-fitting linear combinations to predict simulated $\LXextended$ values listed below the $x$-axis.  The power law exponents for $f_{\rm CGM}$ range between $1.55-2.01$, which are greater than that for $M_{200}$ that range between $1.20-1.63$.  This demonstrates that extended X-ray emission is well-described as a strong function of both variables with $f_{\rm CGM}$ having a somewhat greater effect on average.   

\subsection{SMBH feedback can unbind gaseous halos}   

D20 showed that the central SMBH injects enough feedback energy over its integrated lifetime to unbind a significant fraction of the CGM gas in both simulations.  In the EAGLE low-mass bin, low-sSFR galaxies have substantially higher SMBH masses (median $M_{\rm SMBH}=10^{7.6}\ \msolar$) than their high-sSFR counterparts ($M_{\rm SMBH}=10^{6.8}\ \msolar$), which was shown by \citet{davies19} to be the signpost of AGN feedback expelling CGM gas and curtailing $z=0$ star formation.  \citet{opp20} showed that the expulsion of CGM is a direct result of AGN feedback occurring primarily at $z>1$ in EAGLE, which results in lower $f_{\rm CGM}$ values for $z=0$ low-sSFR, redder galaxies.  EAGLE uses a thermal AGN feedback model \citep{booth09} that applies a single accreted rest mass-to-energy efficiency.

% XXX Rob have deleted references to "radio" mode because Annalisa has explicitly (and reasonably) asked us to avoid this nomenclature.  BDO- Joop had put it in :), agreed. ROB - Throwing Joop under the bus, real Mourinho move... ;-)
TNG also shows a strong anti-correlation between $M_{\rm SMBH}$ and $f_{\rm CGM}$, which also arises from the regulation of star formation via the SMBH expulsion of CGM gas \citep[][D20]{terrazas19}.  However, the corresponding median BH masses for the high and low-sSFR low-mass TNG samples are $M_{\rm SMBH}=10^{8.1}$ and $10^{8.3}\ \msolar$ respectively.  This smaller $M_{\rm SMBH}$ spread belies the much larger difference in $f_{\rm CGM}$, and arises because the kinetic mode AGN feedback, which operates when the central BH reaches a mass of $M_{\rm SMBH}\approx 10^{8}\ \msolar$, is far more efficient at ejecting halo gas than the TNG thermal mode \citep{weinberger17}.  The small scatter and high values in TNG $M_{\rm SMBH}$ at $M_{200} \la 10^{12}\ \msolar$ appear difficult to reconcile with observations \citep[e.g.][]{li19}

\subsection{Is the CGM dominated by cool or hot baryons?}

COS-Halos UV detections of the inner, cool CGM indicate that metal-enriched gas at $T\approx 1-2\times 10^4$ K traces an average $\nh \approx 10^{-3.1}~\cmc$ at $r=20-50$ kpc from $M_*=10^{10.2-11.2}\ \msolar$ galaxies \citep{prochaska17}.  Assuming pressure equilibrium with a $T\ga 10^6$ K halo, the hot gas density at the same radii would be $\nh \la 10^{-5}~\cmc$ \citep{werk14}.  These hot halo densities are at least $1$ dex lower than is predicted by both EAGLE and TNG at $r<50$ kpc, and would produce a hot CGM of much lower luminosity.  Combined with the \citet{prochaska17} calculation that most baryons in $L^*$ halos are accounted for in the cool CGM, we must consider the possibility that the X-ray CGM could be dimmer than these simulations predict.  Therefore, our proposed {\it eROSITA} stacking experiment provides a crucial constraint on the physical nature of the hot halo, which must dominate the CGM volume according to the low filling factor of cool absorbers \citep{stocke13} but not necessarily the mass.  The over-predictions of existing X-ray emission measurements discussed above may point to greater cool baryon fractions than in either simulation.  

\section{Summary} \label{sec:summary} 

We develop a forward modeling pipeline that produces mock {\it eROSITA} stacked observations of X-ray emission from halos ($r>10$ kpc, $\nh<0.22\ \cmc$) around central galaxies using the EAGLE and Illustris-TNG cosmological hydrodynamical simulations.  Both simulations predict that the {\it eROSITA} 4-year all-sky survey, eRASS:8, will result in the robust detection of extended, soft X-ray emission from the hot CGM in stacking analyses.  Our main results are as follows:  

\begin{enumerate}

    \item X-ray halos hosting high-sSFR galaxies with $M_* \approx 10^{10.2-10.8}\ \msolar$ should be detectable out to $30-50$ kpc and be brighter than for low-sSFR galaxies at fixed $M_*$.  Emission around more massive galaxies, $M_* \approx10^{10.7-11.3}\ \msolar$, should be detectable out to $150-200$ kpc.  
    \item TNG predicts a greater dichotomy between high- and low-sSFR X-ray halos at low mass than EAGLE. This is driven by a greater proportion of baryons being retained by star-forming TNG halos.  EAGLE predicts brighter low-sSFR halos than TNG, driven by greater baryon fractions in low-sSFR EAGLE halos.  TNG predicts $3\times$ brighter high-sSFR halos than EAGLE.  
    
    \item Stacked X-ray luminosities are dominated by the brightest halos, more so for low-sSFR than high-sSFR galaxies at low mass.  Deeper {\it eROSITA} surveying at the ecliptic poles should allow individual detections of the brightest halos and constrain the distribution of X-ray halo luminosities contributing to stacks.  

    \item X-ray halos are sensitive probes of the baryon cycle that fuels star-formation and is disrupted by feedback, especially from SMBHs.  X-ray surface brightness distributions should indicate whether the current generation of simulations ejects a sufficient fraction of the CGM, and even help to differentiate between the markedly different implementations of SMBH feedback employed by EAGLE and TNG.
    
\end{enumerate}

Stacking {\it eROSITA} observations will probe galaxies at a variety of distances, and better signal-to-noise will be achieved by stacks of $>10^4$ galaxies out to $z\approx 0.05$.  Additionally, using spectral signatures ({\it eROSITA} has better than 0.1 keV resolution) to separate diffuse gas emission from background contaminants, and measure temperature and metallicity should be possible.  Therefore, our proposed experiment presented here may represent the lowest hanging fruit for CGM science that {\it eROSITA} can achieve.  

\acknowledgments

We thank Urmila Chadayammuri, Dominique Eckert, Ana-Roxana Pop, and Alexey Vikhlinin for useful conversations that added to this work.  We appreciate the anonymous referee for their constructive recommendations.  BDO, AB, WRF, CJ, and RPK acknowledge support from the Smithsonian Institution.  RAC is a Royal Society University Research Fellow.  RPK, WRF, and CJ acknowledge support from the High Resolution Camera program, part of the Chandra X-ray Observatory Center, which is operated by the Smithsonian Astrophysical Observatory for and on behalf of the National Aeronautics Space Administration under contract NAS8-03060.  The study used high performance computing facilities at Liverpool John Moores University, partly funded by the Royal Society and LJMU’s Faculty of Engineering and Technology.  

\bibliography{eROSITA_EAGLE_TNG}{}

\begin{thebibliography}{}
\expandafter\ifx\csname natexlab\endcsname\relax\def\natexlab#1{#1}\fi
\providecommand{\url}[1]{\href{#1}{#1}}
\providecommand{\dodoi}[1]{doi:~\href{http://doi.org/#1}{\nolinkurl{#1}}}
\providecommand{\doeprint}[1]{\href{http://ascl.net/#1}{\nolinkurl{http://ascl.net/#1}}}
\providecommand{\doarXiv}[1]{\href{https://arxiv.org/abs/#1}{\nolinkurl{https://arxiv.org/abs/#1}}}

\bibitem[{{Aguirre} {et~al.}(2001){Aguirre}, {Hernquist}, {Schaye}, {Katz},
  {Weinberg}, \& {Gardner}}]{aguirre01}
{Aguirre}, A., {Hernquist}, L., {Schaye}, J., {et~al.} 2001, \apj, 561, 521,
  \dodoi{10.1086/323370}

\bibitem[{{Anderson} {et~al.}(2016){Anderson}, {Churazov}, \&
  {Bregman}}]{anderson16}
{Anderson}, M.~E., {Churazov}, E., \& {Bregman}, J.~N. 2016, \mnras, 455, 227,
  \dodoi{10.1093/mnras/stv2314}

\bibitem[{{Anderson} {et~al.}(2015){Anderson}, {Gaspari}, {White}, {Wang}, \&
  {Dai}}]{anderson15}
{Anderson}, M.~E., {Gaspari}, M., {White}, S. D.~M., {Wang}, W., \& {Dai}, X.
  2015, \mnras, 449, 3806, \dodoi{10.1093/mnras/stv437}

\bibitem[{{Benson} {et~al.}(2000){Benson}, {Bower}, {Frenk}, \&
  {White}}]{benson00}
{Benson}, A.~J., {Bower}, R.~G., {Frenk}, C.~S., \& {White}, S.~D.~M. 2000,
  \mnras, 314, 557, \dodoi{10.1046/j.1365-8711.2000.03362.x}

\bibitem[{{Bogd{\'a}n} {et~al.}(2017){Bogd{\'a}n}, {Bourdin}, {Forman},
  {Kraft}, {Vogelsberger}, {Hernquist}, \& {Springel}}]{bogdan17}
{Bogd{\'a}n}, {\'A}., {Bourdin}, H., {Forman}, W.~R., {et~al.} 2017, \apj, 850,
  98, \dodoi{10.3847/1538-4357/aa9523}

\bibitem[{{Bogd{\'a}n} {et~al.}(2013{\natexlab{a}}){Bogd{\'a}n}, {Forman},
  {Kraft}, \& {Jones}}]{bogdan13a}
{Bogd{\'a}n}, {\'A}., {Forman}, W.~R., {Kraft}, R.~P., \& {Jones}, C.
  2013{\natexlab{a}}, \apj, 772, 98, \dodoi{10.1088/0004-637X/772/2/98}

\bibitem[{{Bogd{\'a}n} {et~al.}(2013{\natexlab{b}}){Bogd{\'a}n}, {Forman},
  {Vogelsberger}, {Bourdin}, {Sijacki}, {Mazzotta}, {Kraft}, {Jones},
  {Gilfanov}, {Churazov}, \& {David}}]{bogdan13b}
{Bogd{\'a}n}, {\'A}., {Forman}, W.~R., {Vogelsberger}, M., {et~al.}
  2013{\natexlab{b}}, \apj, 772, 97, \dodoi{10.1088/0004-637X/772/2/97}

\bibitem[{{Booth} \& {Schaye}(2009)}]{booth09}
{Booth}, C.~M., \& {Schaye}, J. 2009, \mnras, 398, 53,
  \dodoi{10.1111/j.1365-2966.2009.15043.x}

\bibitem[{{Crain} {et~al.}(2010){Crain}, {McCarthy}, {Frenk}, {Theuns}, \&
  {Schaye}}]{crain10}
{Crain}, R.~A., {McCarthy}, I.~G., {Frenk}, C.~S., {Theuns}, T., \& {Schaye},
  J. 2010, \mnras, 407, 1403, \dodoi{10.1111/j.1365-2966.2010.16985.x}

\bibitem[{{Crain} {et~al.}(2013){Crain}, {McCarthy}, {Schaye}, {Theuns}, \&
  {Frenk}}]{crain13}
{Crain}, R.~A., {McCarthy}, I.~G., {Schaye}, J., {Theuns}, T., \& {Frenk},
  C.~S. 2013, \mnras, 432, 3005, \dodoi{10.1093/mnras/stt649}

\bibitem[{{Crain} {et~al.}(2015){Crain}, {Schaye}, {Bower}, {Furlong},
  {Schaller}, {Theuns}, {Dalla Vecchia}, {Frenk}, {McCarthy}, {Helly},
  {Jenkins}, {Rosas-Guevara}, {White}, \& {Trayford}}]{crain15}
{Crain}, R.~A., {Schaye}, J., {Bower}, R.~G., {et~al.} 2015, \mnras, 450, 1937,
  \dodoi{10.1093/mnras/stv725}

\bibitem[{{Dauser} {et~al.}(2019){Dauser}, {Falkner}, {Lorenz}, {Kirsch},
  {Peille}, {Cucchetti}, {Schmid}, {Brand}, {Oertel}, {Smith}, \&
  {Wilms}}]{sixte}
{Dauser}, T., {Falkner}, S., {Lorenz}, M., {et~al.} 2019, \aap, 630, A66,
  \dodoi{10.1051/0004-6361/201935978}

\bibitem[{{Davies} {et~al.}(2019){Davies}, {Crain}, {McCarthy}, {Oppenheimer},
  {Schaye}, {Schaller}, \& {McAlpine}}]{davies19}
{Davies}, J.~J., {Crain}, R.~A., {McCarthy}, I.~G., {et~al.} 2019, \mnras, 485,
  3783, \dodoi{10.1093/mnras/stz635}

\bibitem[{{Davies} {et~al.}(2020){Davies}, {Crain}, {Oppenheimer}, \&
  {Schaye}}]{davies20}
{Davies}, J.~J., {Crain}, R.~A., {Oppenheimer}, B.~D., \& {Schaye}, J. 2020,
  \mnras, 491, 4462, \dodoi{10.1093/mnras/stz3201}

\bibitem[{{Ford} {et~al.}(2014){Ford}, {Dav{\'e}}, {Oppenheimer}, {Katz},
  {Kollmeier}, {Thompson}, \& {Weinberg}}]{ford14}
{Ford}, A.~B., {Dav{\'e}}, R., {Oppenheimer}, B.~D., {et~al.} 2014, \mnras,
  444, 1260, \dodoi{10.1093/mnras/stu1418}

\bibitem[{{Ford} {et~al.}(2013){Ford}, {Oppenheimer}, {Dav{\'e}}, {Katz},
  {Kollmeier}, \& {Weinberg}}]{ford13}
{Ford}, A.~B., {Oppenheimer}, B.~D., {Dav{\'e}}, R., {et~al.} 2013, \mnras,
  432, 89, \dodoi{10.1093/mnras/stt393}

\bibitem[{{Forman} {et~al.}(1985){Forman}, {Jones}, \& {Tucker}}]{forman85}
{Forman}, W., {Jones}, C., \& {Tucker}, W. 1985, \apj, 293, 102,
  \dodoi{10.1086/163218}

\bibitem[{{Fruscione} {et~al.}(2006){Fruscione}, {McDowell}, {Allen},
  {Brickhouse}, {Burke}, {Davis}, {Durham}, {Elvis}, {Galle}, {Harris},
  {Huenemoerder}, {Houck}, {Ishibashi}, {Karovska}, {Nicastro}, {Noble},
  {Nowak}, {Primini}, {Siemiginowska}, {Smith}, \& {Wise}}]{ciao}
{Fruscione}, A., {McDowell}, J.~C., {Allen}, G.~E., {et~al.} 2006, Society of
  Photo-Optical Instrumentation Engineers (SPIE) Conference Series, Vol. 6270,
  {CIAO: Chandra's data analysis system}, 62701V, \dodoi{10.1117/12.671760}

\bibitem[{{Goulding} {et~al.}(2016){Goulding}, {Greene}, {Ma}, {Veale},
  {Bogdan}, {Nyland}, {Blakeslee}, {McConnell}, \& {Thomas}}]{goulding16}
{Goulding}, A.~D., {Greene}, J.~E., {Ma}, C.-P., {et~al.} 2016, \apj, 826, 167,
  \dodoi{10.3847/0004-637X/826/2/167}

\bibitem[{{Li} {et~al.}(2017){Li}, {Bregman}, {Wang}, {Crain}, {Anderson}, \&
  {Zhang}}]{li17}
{Li}, J.-T., {Bregman}, J.~N., {Wang}, Q.~D., {et~al.} 2017, \apjs, 233, 20,
  \dodoi{10.3847/1538-4365/aa96fc}

\bibitem[{{Li} \& {Wang}(2013)}]{li13}
{Li}, J.-T., \& {Wang}, Q.~D. 2013, \mnras, 428, 2085,
  \dodoi{10.1093/mnras/sts183}

\bibitem[{{Li} {et~al.}(2019){Li}, {Habouzit}, {Genel}, {Somerville},
  {Terrazas}, {Bell}, {Pillepich}, {Nelson}, {Weinberger}, {Rodriguez-Gomez},
  {Ma}, {Pakmor}, {Hernquist}, \& {Vogelsberger}}]{li19}
{Li}, Y., {Habouzit}, M., {Genel}, S., {et~al.} 2019, arXiv e-prints,
  arXiv:1910.00017.
\newblock \doarXiv{1910.00017}

\bibitem[{{Li} {et~al.}(2006){Li}, {Wang}, {Irwin}, \& {Chaves}}]{li06}
{Li}, Z., {Wang}, Q.~D., {Irwin}, J.~A., \& {Chaves}, T. 2006, \mnras, 371,
  147, \dodoi{10.1111/j.1365-2966.2006.10682.x}

\bibitem[{{Liang} \& {Chen}(2014)}]{liang14}
{Liang}, C.~J., \& {Chen}, H.-W. 2014, \mnras, 445, 2061,
  \dodoi{10.1093/mnras/stu1901}

\bibitem[{{McAlpine} {et~al.}(2016){McAlpine}, {Helly}, {Schaller}, {Trayford},
  {Qu}, {Furlong}, {Bower}, {Crain}, {Schaye}, {Theuns}, {Dalla Vecchia},
  {Frenk}, {McCarthy}, {Jenkins}, {Rosas-Guevara}, {White}, {Baes}, {Camps}, \&
  {Lemson}}]{mcalpine16}
{McAlpine}, S., {Helly}, J.~C., {Schaller}, M., {et~al.} 2016, Astronomy and
  Computing, 15, 72, \dodoi{10.1016/j.ascom.2016.02.004}

\bibitem[{{Merloni} {et~al.}(2012){Merloni}, {Predehl}, {Becker},
  {B{\"o}hringer}, {Boller}, {Brunner}, {Brusa}, {Dennerl}, {Freyberg},
  {Friedrich}, {Georgakakis}, {Haberl}, {Hasinger}, {Meidinger}, {Mohr},
  {Nandra}, {Rau}, {Reiprich}, {Robrade}, {Salvato}, {Santangelo}, {Sasaki},
  {Schwope}, {Wilms}, \& {German eROSITA Consortium}}]{erosita}
{Merloni}, A., {Predehl}, P., {Becker}, W., {et~al.} 2012, ArXiv e-prints.
\newblock \doarXiv{1209.3114}

\bibitem[{{Nelson} {et~al.}(2018{\natexlab{a}}){Nelson}, {Pillepich},
  {Springel}, {Weinberger}, {Hernquist}, {Pakmor}, {Genel}, {Torrey},
  {Vogelsberger}, {Kauffmann}, {Marinacci}, \& {Naiman}}]{nelson18a}
{Nelson}, D., {Pillepich}, A., {Springel}, V., {et~al.} 2018{\natexlab{a}},
  \mnras, 475, 624, \dodoi{10.1093/mnras/stx3040}

\bibitem[{{Nelson} {et~al.}(2018{\natexlab{b}}){Nelson}, {Kauffmann},
  {Pillepich}, {Genel}, {Springel}, {Pakmor}, {Hernquist}, {Weinberger},
  {Torrey}, {Vogelsberger}, \& {Marinacci}}]{nelson18b}
{Nelson}, D., {Kauffmann}, G., {Pillepich}, A., {et~al.} 2018{\natexlab{b}},
  \mnras, 477, 450, \dodoi{10.1093/mnras/sty656}

\bibitem[{{Oppenheimer} {et~al.}(2018){Oppenheimer}, {Schaye}, {Crain}, {Werk},
  \& {Richings}}]{opp18c}
{Oppenheimer}, B.~D., {Schaye}, J., {Crain}, R.~A., {Werk}, J.~K., \&
  {Richings}, A.~J. 2018, \mnras, 481, 835, \dodoi{10.1093/mnras/sty2281}

\bibitem[{{Oppenheimer} {et~al.}(2016){Oppenheimer}, {Crain}, {Schaye},
  {Rahmati}, {Richings}, {Trayford}, {Tumlinson}, {Bower}, {Schaller}, \&
  {Theuns}}]{opp16}
{Oppenheimer}, B.~D., {Crain}, R.~A., {Schaye}, J., {et~al.} 2016, \mnras, 460,
  2157, \dodoi{10.1093/mnras/stw1066}

\bibitem[{{Oppenheimer} {et~al.}(2020){Oppenheimer}, {Davies}, {Crain},
  {Wijers}, {Schaye}, {Werk}, {Burchett}, {Trayford}, \& {Horton}}]{opp20}
{Oppenheimer}, B.~D., {Davies}, J.~J., {Crain}, R.~A., {et~al.} 2020, \mnras,
  491, 2939, \dodoi{10.1093/mnras/stz3124}

\bibitem[{{O'Sullivan} {et~al.}(2001){O'Sullivan}, {Forbes}, \&
  {Ponman}}]{osullivan01}
{O'Sullivan}, E., {Forbes}, D.~A., \& {Ponman}, T.~J. 2001, \mnras, 328, 461,
  \dodoi{10.1046/j.1365-8711.2001.04890.x}

\bibitem[{{Pillepich} {et~al.}(2018){Pillepich}, {Springel}, {Nelson}, {Genel},
  {Naiman}, {Pakmor}, {Hernquist}, {Torrey}, {Vogelsberger}, {Weinberger}, \&
  {Marinacci}}]{pillepich18}
{Pillepich}, A., {Springel}, V., {Nelson}, D., {et~al.} 2018, \mnras, 473,
  4077, \dodoi{10.1093/mnras/stx2656}

\bibitem[{{Prochaska} {et~al.}(2017){Prochaska}, {Werk}, {Worseck}, {Tripp},
  {Tumlinson}, {Burchett}, {Fox}, {Fumagalli}, {Lehner}, {Peeples}, \&
  {Tejos}}]{prochaska17}
{Prochaska}, J.~X., {Werk}, J.~K., {Worseck}, G., {et~al.} 2017, \apj, 837,
  169, \dodoi{10.3847/1538-4357/aa6007}

\bibitem[{{Rahmati} {et~al.}(2016){Rahmati}, {Schaye}, {Crain}, {Oppenheimer},
  {Schaller}, \& {Theuns}}]{rahmati16}
{Rahmati}, A., {Schaye}, J., {Crain}, R.~A., {et~al.} 2016, \mnras, 459, 310,
  \dodoi{10.1093/mnras/stw453}

\bibitem[{{Schaye} {et~al.}(2015){Schaye}, {Crain}, {Bower}, {Furlong},
  {Schaller}, {Theuns}, {Dalla Vecchia}, {Frenk}, {McCarthy}, {Helly},
  {Jenkins}, {Rosas-Guevara}, {White}, {Baes}, {Booth}, {Camps}, {Navarro},
  {Qu}, {Rahmati}, {Sawala}, {Thomas}, \& {Trayford}}]{schaye15}
{Schaye}, J., {Crain}, R.~A., {Bower}, R.~G., {et~al.} 2015, \mnras, 446, 521,
  \dodoi{10.1093/mnras/stu2058}

\bibitem[{{Smith} {et~al.}(2001){Smith}, {Brickhouse}, {Liedahl}, \&
  {Raymond}}]{smith01}
{Smith}, R.~K., {Brickhouse}, N.~S., {Liedahl}, D.~A., \& {Raymond}, J.~C.
  2001, \apjl, 556, L91, \dodoi{10.1086/322992}

\bibitem[{{Spitzer}(1956)}]{spitzer56}
{Spitzer}, Lyman, J. 1956, \apj, 124, 20, \dodoi{10.1086/146200}

\bibitem[{{Springel}(2005)}]{springel05}
{Springel}, V. 2005, \mnras, 364, 1105,
  \dodoi{10.1111/j.1365-2966.2005.09655.x}

\bibitem[{{Springel}(2010)}]{springel10}
---. 2010, \mnras, 401, 791, \dodoi{10.1111/j.1365-2966.2009.15715.x}

\bibitem[{{Stocke} {et~al.}(2013){Stocke}, {Keeney}, {Danforth}, {Shull},
  {Froning}, {Green}, {Penton}, \& {Savage}}]{stocke13}
{Stocke}, J.~T., {Keeney}, B.~A., {Danforth}, C.~W., {et~al.} 2013, \apj, 763,
  148, \dodoi{10.1088/0004-637X/763/2/148}

\bibitem[{{Terrazas} {et~al.}(2019){Terrazas}, {Bell}, {Pillepich}, {Nelson},
  {Somerville}, {Genel}, {Weinberger}, {Habouzit}, {Li}, {Hernquist}, \&
  {Vogelsberger}}]{terrazas19}
{Terrazas}, B.~A., {Bell}, E.~F., {Pillepich}, A., {et~al.} 2019, arXiv
  e-prints, arXiv:1906.02747.
\newblock \doarXiv{1906.02747}

\bibitem[{{Truong} {et~al.}(2020){Truong}, {Pillepich}, {Werner}, {Nelson},
  {Lakhchaura}, {Weinberger}, {Springel}, {Vogelsberger}, \&
  {Hernquist}}]{truong20}
{Truong}, N., {Pillepich}, A., {Werner}, N., {et~al.} 2020, \mnras,
  \dodoi{10.1093/mnras/staa685}

\bibitem[{{Tumlinson} {et~al.}(2011){Tumlinson}, {Thom}, {Werk}, {Prochaska},
  {Tripp}, {Weinberg}, {Peeples}, {O'Meara}, {Oppenheimer}, {Meiring}, {Katz},
  {Dav{\'e}}, {Ford}, \& {Sembach}}]{tumlinson11}
{Tumlinson}, J., {Thom}, C., {Werk}, J.~K., {et~al.} 2011, Science, 334, 948,
  \dodoi{10.1126/science.1209840}

\bibitem[{{Turner} {et~al.}(2014){Turner}, {Schaye}, {Steidel}, {Rudie}, \&
  {Strom}}]{turner14}
{Turner}, M.~L., {Schaye}, J., {Steidel}, C.~C., {Rudie}, G.~C., \& {Strom},
  A.~L. 2014, \mnras, 445, 794, \dodoi{10.1093/mnras/stu1801}

\bibitem[{{Weinberger} {et~al.}(2017){Weinberger}, {Springel}, {Hernquist},
  {Pillepich}, {Marinacci}, {Pakmor}, {Nelson}, {Genel}, {Vogelsberger},
  {Naiman}, \& {Torrey}}]{weinberger17}
{Weinberger}, R., {Springel}, V., {Hernquist}, L., {et~al.} 2017, \mnras, 465,
  3291, \dodoi{10.1093/mnras/stw2944}

\bibitem[{{Werk} {et~al.}(2014){Werk}, {Prochaska}, {Tumlinson}, {Peeples},
  {Tripp}, {Fox}, {Lehner}, {Thom}, {O'Meara}, {Ford}, {Bordoloi}, {Katz},
  {Tejos}, {Oppenheimer}, {Dav{\'e}}, \& {Weinberg}}]{werk14}
{Werk}, J.~K., {Prochaska}, J.~X., {Tumlinson}, J., {et~al.} 2014, \apj, 792,
  8, \dodoi{10.1088/0004-637X/792/1/8}

\bibitem[{{White} \& {Frenk}(1991)}]{white91}
{White}, S.~D.~M., \& {Frenk}, C.~S. 1991, \apj, 379, 52,
  \dodoi{10.1086/170483}

\bibitem[{{White} \& {Rees}(1978)}]{white78}
{White}, S.~D.~M., \& {Rees}, M.~J. 1978, \mnras, 183, 341,
  \dodoi{10.1093/mnras/183.3.341}

\bibitem[{{ZuHone} \& {Hallman}(2016)}]{zuhone16}
{ZuHone}, J.~A., \& {Hallman}, E.~J. 2016, {pyXSIM: Synthetic X-ray
  observations generator}.
\newblock \doeprint{1608.002}

\end{thebibliography}
\bibliographystyle{aasjournal}

\end{document}